\documentclass[aps,prc,showpacs,floatfix,twocolumn]{revtex4-1}
\usepackage{graphicx}
\usepackage{amsmath}
\usepackage{subcaption}
\usepackage{subfig}
\begin{document}
\title{Radiative proton capture cross sections in the mass range $40-54$}
\author{Dipti Chakraborty}
\email{diptichakraborty2011@gmail.com}
\author{Saumi Dutta}
\email{saumidutta89@gmail.com}
\author{G. Gangopadhyay}
\email{ggphy@caluniv.ac.in}
\author{Abhijit Bhattacharyya}
\email{abhattacharyyacu@gamil.com}
\affiliation{Department of Physics, University of Calcutta, 
92 Acharya Prafulla Chandra Road, Kolkata-700009}

\begin{abstract}
Proton capture  cross sections in the energy range of astrophysical interest 
for mass region 40-54 have been calculated in
the Hauser-Feshbach formalism with reaction code TALYS1.6. 
The density dependent M3Y effective 
nucleon-nucleon interaction folded with target radial matter densities from 
relativistic mean field approach is used
to obtain the semimicroscopic optical potential. A  definite normalization of 
potential well depths has been used  over the entire mass region.
The $(p,\gamma)$ rates of some reactions, important in the astrophysical scenario, are calculated using the potential
in the relevant mass region.
\end{abstract}
\maketitle

\section{Introduction}

During stellar burning, reactions involving different seed nuclei assume 
importance at different
temperature and burning zones. The seeds in the concerned mass range A=40-54 
are mainly produced during hydrostatic carbon burning and explosive oxygen 
burning. These then take part in synthesizing more massive elements via various 
reactions
occurring in later phases of evolution~\cite{seed_reaction1,seed_reaction2}. 

The principle energy generating processes in stars i.e the  pp-cycle, CNO cycle,
HCNO cycle, {\em rp} process etc are reactions which require
continuous addition of protons against the Coulomb barrier. Certain 
astrophysical sites, such as X-ray bursters, involve high flux of protons 
at temperature $\sim$ several GK and matter density $\sim$ $10^{6}$ g/cc 
that initiate a rapid proton capture process which ultimately results in 
a thermal burst of very short duration, peaked in X ray regime. Knowledge of
the cross sections, and equivalently, 
the rates of the reactions occurring in these sites are required to study 
complete nucleosynthesis via a network calculation. However, it is difficult to 
measure all the essential rates in terrestrial laboratories due to 
unavailability and instability of required targets. Reaction rates calculated 
in a theoretical approach may provide necessary information to this context 
after proper validation of theory with the experimental data. For abundance calculation in explosive astrophysical environment, the concerned network must have to take various quantities like temperature, pressure, proton mass fraction along with forward and reverse reaction rates into account. So, we need to take care about the proper tuning of the interaction potential.  

Many works have been devoted so far to study the theoretical capture cross sections by constructing different nucleon-nucleus potentials. Rauscher {\em et. al.}~\cite {raus1,raus2} have calculated reaction rates in a global approach and suggested that statistical model calculation can be made improved using locally tuned nuclear properties like optical potential. Reaction rates from phenomenological approach i.e, with phenomenological global optical potentials or even with semimicroscopic optical potential with phenomenological densities give rise to uncertainty and the reaction rates have to be varied by large factors to study their effects. Prediction of rates, as has been done in Ref.~\cite{schatz1}, gave rise to uncertainties away from the stability valley. Phenomenological global optical potential should not be expected to provide an adequate description of nucleon-nucleus interaction as differences in nuclear structure among adjacent nuclei do not allow simple and smooth Z- and A- dependence of the Woods-Saxon parameters. Microscopic optical potentials obtained by folding with appropriate microscopic densities are expected to be more accurate and do not require frequent variation of the reaction rates. On the other hand, calculation of reaction rates in a microscopic or semimicroscopic with microscopic density prescription approach is far more free from these uncertainties. Recently, 
the semimicroscopic optical potential obtained in a folding model prescription 
has proved to be highly successful in explaining various nuclear phenomena.
For example, Bauge  {\em et al.}~\cite{bauge1,bauge2} have constructed a 
lane-consistent semimicroscopic optical potential to study elastic scattering and 
differential and total  cross sections for nuclei over a 
broad mass range. The theoretical cross section calculation requires a complete 
knowledge of various ingredients such as transmission coefficients, i.e., 
transition probabilities (averaged over resonances of the compound nucleus 
formed upon radiative proton captures) between various states which in turn 
depend on the level schemes, lifetimes of the states, level densities, 
$\gamma$ ray strength functions, nuclear masses, giant dipole resonance 
parameters, etc. 

Application of statistical model requires sufficiently high nuclear level 
density in the compound nuclear state. The theoretical reactions
are generally derived in Hauser-Feshabach (HF) formalism which assumes presence of 
sufficiently large number of resonances at relevant energies.  
The cross sections are in general sum of contribution from different reaction 
mechanisms depending on projectile energies. At higher energies, presence of 
many close and overlapping resonances 
allow one to calculate an average cross sections using HF approach. 
Sometimes there are interference effects between single resonance and 
direct capture. 
The nonresonant reaction cross sections are mainly determined by the direct 
capture
 transitions to the ground states and  low excited states.
In light nuclei and low energy regime, level densities are generally low, especially for targets 
near closed shells with widely spaced nuclear levels and close to drip lines 
with low particle separation energies and Q-values. Hence, application of 
statistical model to these nuclei at astrophysical temperatures is somewhat 
problematic and requires careful study.

Optical potential is a very important ingredient in HF statistical model calculation. Here, we have constructed an optical potential by folding the density dependent M3Y 
(DDM3Y) interaction with the densities of the target. 
The DDM3Y interaction has proved to be successful in explaining various nuclear properties.
 For example, folded DDM3Y nucleon-nucleus interaction potential has been
successfully  used to study the incompressibility of infinite nuclear matter \cite{dnbasu1} 
and
 radioactivity lifetimes of spherical proton rich nuclei \cite{dnbasu2}.

The paper is organized as follows. In the next section we have presented the 
framework of our calculation. In section III we discuss the results and in the last section we summarize.

\section{Model Calculation}

Proton capture reactions for nuclei A $\geq$ 40 have Q-values $\geq$ 5 MeV which
increase with increasing number of neutrons. We have studied proton capture reactions 
in several nuclei in the mass range $A=40-54$ in semimicroscopic HF formalism and 
compared our results 
with experiments. All these nuclei have Q values for proton capture at ground 
states within $\sim$ 5 MeV to 10 MeV.

We have calculated the reaction cross sections in HF approach using 
the folding model formalism with the TALYS1.6 code~\cite{talys1.6}. The code TALYS incorporates much more physics than the other codes available previously. It takes care of preequlibrium emission, detailed competition between all open channels, detailed width fluctuation correction, coherent inclusion of fission channel, coupled channel calculaion for deformed nuclei, multiparticle emission etc. Details can be found in Talys1.6 manual. Along with these advantages of the code, the code can be used with fully microscopic inputs, and hence is suitable for determination of unknown rates, a feature utilized in present calculation. The present approach scores out the phenomenological method of Goriely {\em et. al.}~\cite{gori2008} with the same code. The HF  formalism generally considers the 
formation of a compound nuclear state. The present method has 
been adopted in a number  of our recent works~\cite{gg, lahiri1, lahiri2,lahiri3, saumi,dipti}. Radial densities have been obtained  from relativistic mean 
field (RMF) approach using the FSUGold Lagrangian density~\cite{fsugold}. Details of which are given in Ref~\cite{bhat1,bhat2}. The target 
is assumed to be spherical and calculations are done in coordinate space. Considering finite size of the proton, 
charge densities are obtained by convoluting the point proton density with a standard Gaussian form factor 
$F({\bf r})$~\cite{gaussformfactor},

\begin{equation}
\rho_{ch}({\bf r})=e \int \rho({\bf r^\prime})F({\bf r}-{\bf r^\prime}) d{\bf r^\prime}
\end{equation}
where
\begin{equation}
F(r)=(a\sqrt\pi)^{-3}exp(-\frac{r^{2}}{a^{2}})
\end{equation}
Here we have $a$=$\sqrt{2/3}$$a_p$, with $a_p$=0.80 fm being the root mean 
square (rms) charge radius of proton. The effect of centre of mass correction is neglected while calculating the charge density or radius as it goes as A$^{-4/3}$ as given in Ref~\cite{quentin}.

The DDM3Y interaction  is folded with the radial  
 densities of targets obtained from RMF approach. 
This interaction potential is then incorporated into the reaction code.
 The interaction, in MeV, at a distance $r$ (in $fm$)is given by \cite{kobos,ak1,ak2}, 
\begin{equation}
v(r,\rho,E)=t^{M3Y}(r,E)g(\rho)
\end{equation}
with the M3Y interaction (including a zero range pseudo potential) given by \cite{m3y1,m3y2},
\begin{equation}
t^{M3Y}=7999\frac{e^{-4r}}{4r}-2134\frac{e^{-2.5r}}{2.5r}-276(1-0.005\frac{E}{A}) \delta(r)
\end{equation}
E is the energy in MeV in the center of mass frame of the projectile and $\rho$ being the nuclear density. 
The density dependent factor $g(\rho)$ is given as~\cite{cbeta}.
\begin{equation}
g(\rho)=C[1-\beta\rho^{2/3}]
\end{equation}
The constants C and $\beta$ are assigned values 2.07 and 1.624 fm$^{2}$, 
respectively, obtained from nuclear matter calculation~\cite{dnbasu1}. 
We have further included a spin-orbit term coupled with phenomenological energy 
dependent potential depths according to Sheerbaum prescription~\cite{scheerbaum} as
\begin{equation}
U^{so}_{n(p)}(r)=(\lambda_{vso}+i\lambda_{wso})\frac{1}{r}\frac{d}{dr}
\left(\frac{2}{3}\rho_{p(n)}+\frac{1}{3}\rho_{n(p)}\right)
\end{equation}
where,
\begin{equation}
\lambda_{vso}=130\exp(-0.013~E)+40
\end{equation}
\begin{equation}
\lambda_{wso}=-0.2(E-20)
\end{equation}

DDM3Y interaction potential provides only the real part and the imaginary part of the optical model potential has been taken to be identical with the real part. The folded potential is then renormalized to obtain the reasonable agreement 
with experimental data. We have used the same  renormalization factor, 0.9, 
for the entire mass region and have used the same potential for the real and 
the imaginary parts of the potential. The advantage of same renormalization factor for entire mass region 
ensures that the present method can be extended to nuclei where experimental
data are not available.

Nuclear level densities are important ingredients of the calculation~\cite{ld_uncert1,ld_uncert2}. 
We have used Goriely's microscopic results for level density~\cite{ldmodel} 
and Hartree-Fock-Bogoliubov model
for E1 $\gamma$ ray strength function~\cite{gammasf}. These choices were
used in some of our previous calculations also. We always try to avoid too much variation of the theory or input parameters to have easy extrapolation of the rate parameters in future use. Hence, we decided to keep the nuclear level density model same as in our previous publications in the mass range 55-80~\cite{gg, lahiri1, lahiri2,lahiri3, saumi,dipti}. We have checked that this choice works well in the mass region A=$40-80$. Width fluctuation correction, which is due to the retention of some memory
of the initial channel,
is also important. Though it assumes significance mainly in elastic 
scattering, it is also important
near threshold energies, where competition cusps arise due to the existence of 
different channel strengths. 
density profile figure
\begin{figure*}[hbt]
\center
\includegraphics[scale=0.35]{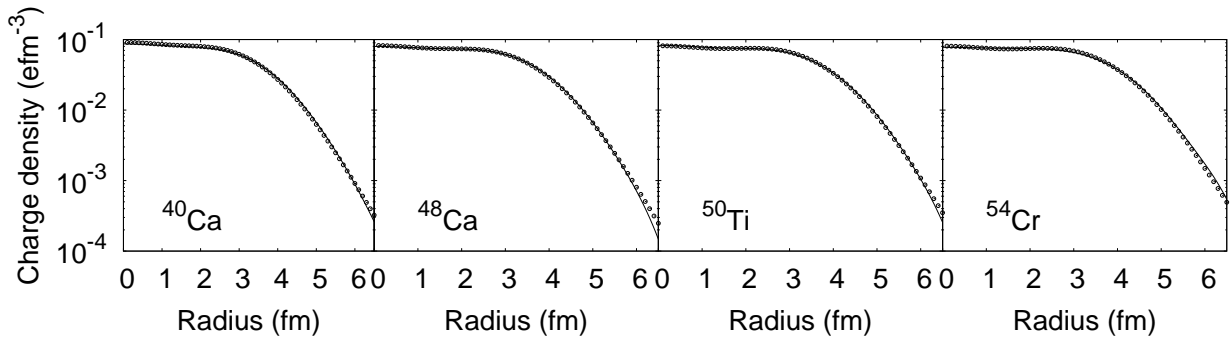}\\
\caption{\label{density_profile}Density profiles of some nuclei in the mass range of interest. Solid lines denote our results and  discrete points indicate results from the Fourier-Bessel [FB] parameterizations obtained from fitting the experimental data.}
\end{figure*}
\begin{table*}[tbh]
\begin{tabular}{ c rrrr| crrrr | crrrr} 
\hline
Nucleus     & \multicolumn{2}{c} {B.E(MeV)}&   \multicolumn{2}{c|} {Charge radius(fm)} &
Nucleus     & \multicolumn{2}{c} {B.E(MeV)}&   \multicolumn{2}{c|} {Charge radius(fm)} &
 Nucleus     & \multicolumn{2}{c} {B.E(MeV)}&   \multicolumn{2}{c} {Charge radius(fm)} 
\\
  &   Theory   &      Expt.  &        Theory   &      Expt.  &
  &   Theory   &      Expt.  &        Theory   &      Expt.  &     
  & Theory   &      Expt.  &        Theory   &      Expt.  \\
\hline
$^{40}$Ar& 340.36 &343.81&  3.36&  3.42&
$^{41}$K& 351.76 &351.62& 3.40& 3.45&
$^{40}$Ca&341.90 &342.05& 3.43&  3.47\\
$^{42}$Ca&360.80&361.89& 3.43& 3.50&
$^{43}$Ca&370.13&369.83& 3.44& 3.49&
$^{44}$Ca&375.80  &380.96&3.44& 3.51\\
$^{46}$Ca&395.85 &398.77& 3.45& 3.49&
$^{45}$Sc&386.85 &387.85& 3.48 & 3.54&
$^{46}$Ti&393.69&398.20 &3.52&3.60\\
$^{47}$Ti&403.93&407.08&3.52&3.59&
$^{48}$Ti&414.02&418.70&3.53&3.59&
$^{49}$Ti&423.59&426.85& 3.53 &3.57\\
$^{50}$Ti&432.34 &437.78&3.53 & 3.57&
$^{51}$V&441.25 &445.85&3.57 & 3.59&
$^{50}$Cr&428.69 &435.05&3.60& 3.66\\
$^{52}$Cr&450.07 &456.35& 3.60& 3.64&
$^{53}$Cr&459.00  &464.29&3.61 & 3.65&
$^{54}$Cr&467.19 &474.01 & 3.63 & 3.68\\
$^{54}$Fe&464.91  &471.76& 3.66& 3.69\\
\hline
\end{tabular}
\caption{\label{be}Comparison of experimental and calculated binding energy (BE) and charge radii
values  for all the stable nuclei in the mass range of interest. Theoretical
BE values are from RMF with the 
N$_{p}$N$_{n}$ correction~\cite{becor,becor1}.}
\end{table*}
 
Nuclear masses, if experimental values are not available, are taken from Ref.~\cite{massmodel}. Preequilibrium effects have been included in cross section 
calculation.  Thirty  discrete levels both for target and residual nuclei have been  
taken into consideration in
Hauser-Feshbach decay and $\gamma$ ray cascade. Full $j,l$ coupling between the states is 
considered.
The TALYS database includes all these necessary ingredients. 

Astrophysical proton capture reactions are mainly important in a definite energy
window, termed as effective Gamow energy window, distributed around the Gamow 
peak~\cite{gamow}. In charged particle reactions, the window arises from the 
folding of the Maxwell-Boltzmann distribution of the particles with the Coulomb barrier 
penetrability. Consequently, rates for charged particle reactions are 
largely suppressed at both low and high energies. Both the peak energy and the 
width depend on charges of the projectile and the target and temperature of
the astrophysical environment.

Proton capture reaction cross sections have been calculated for energies 
relevant to this energy window and compared with available experimental data. The relevant gamow energy window for this mass region lies in between 1-2 MeV corresponding to the temperature~$\sim$2GK.
S factors are commonly used in astrophysical applications at such low 
incident energies where cross sections show strong energy dependence. S factors remove the Coulomb dependence
and hence it is a slowly varying function of energy. It is given by, 
\begin{equation} S(E)=E~\sigma(E)~exp(2\pi \eta)
\end{equation}
Here $\eta$ is known as Sommerfeld parameter given by,
\begin{equation}
\eta=0.989534~Z_{p}~Z_{t}~\mu/~E. 
\end{equation}

 with $\mu$=$\frac{m_{p}m_{t}}{m_{p}+m_{t}}$ being the reduced mass, $Z_{p}$, $Z_{t}$ are proton numbers of projectile and target, respectively, and $m_{p}$, $m_{t}$ are masses of projectile and target, respectively.

Finally, we have calculated the  rates of some reactions which have been 
identified as important by Parikh {\em et al.}~\cite{parikh} in XRB  nucleosynthesis. 
They have done post-processing calculations for type I X-ray burst using ten 
different models with different temperature density profiles. Three of them 
are from Koike {\em et al.}~\cite{koike}, Schatz {\em et al.}~\cite{schatz}, and Fisker {\em et al.}~\cite{fisker}. They have further parameterized
the model of Koike {\em et al.} to probe the effects of burst peak temperature, burst 
duration, and metallicities. Using these ten models, they have studied the
impacts of thermonuclear reaction rates on XRB yields. They have identified some reactions as important in the mass range of our interest. We have calculated the rates of those 
reactions with our theoretical model and further compared with the rates predicted by NON-SMOKER reaction code~\cite{nonsmoker}.

\section{Results} 
\subsection {Results of RMF calculations}
In the present work, RMF approach is used to obtain density distributions 
which are then folded 
with the interaction to obtain the optical potential. 
We have plotted the 
charge density profiles of some nuclei in the relevant mass region in 
Fig. \ref{density_profile} and compared with experimental values extracted from
Fourier-Bessel parameterizations~\cite{fourbessel}. 
One can see that our theory can reproduce the experimental density very well.
Further, in Table \ref{be} we have listed the calculated binding energy and charge radii values 
with the experimental measurements from Ref~\cite{binding_energy} and Ref~\cite{rmscharge} respectively. In case of binding energies the theoretical values have been corrected following the prescription~\cite{becor,becor1}.
 One can readily see that the agreement between theory and measurement is reasonable with the difference between them is less than 1.5\%. In case of charge radii the difference between theory and experiment is less than 2.1\% as can be seen from Table \ref{be}.

\subsection{Cross section Calculation and Astrophysical S factor}

\begin{figure}[hbt]
\center
\begin{tabular}{@{}cc@{}}
\includegraphics[width=.25\textwidth]{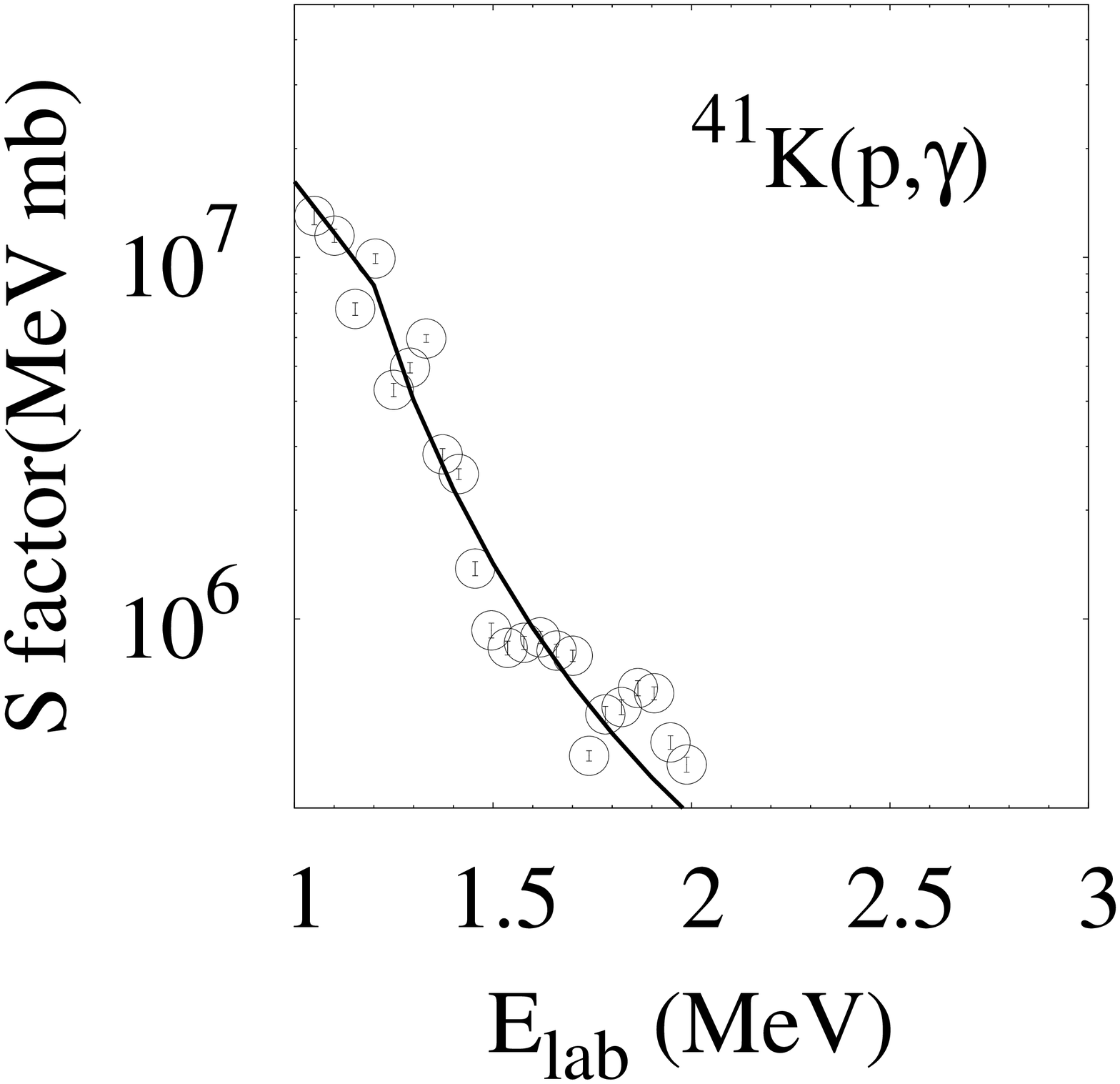} &
\includegraphics[width=.25\textwidth]{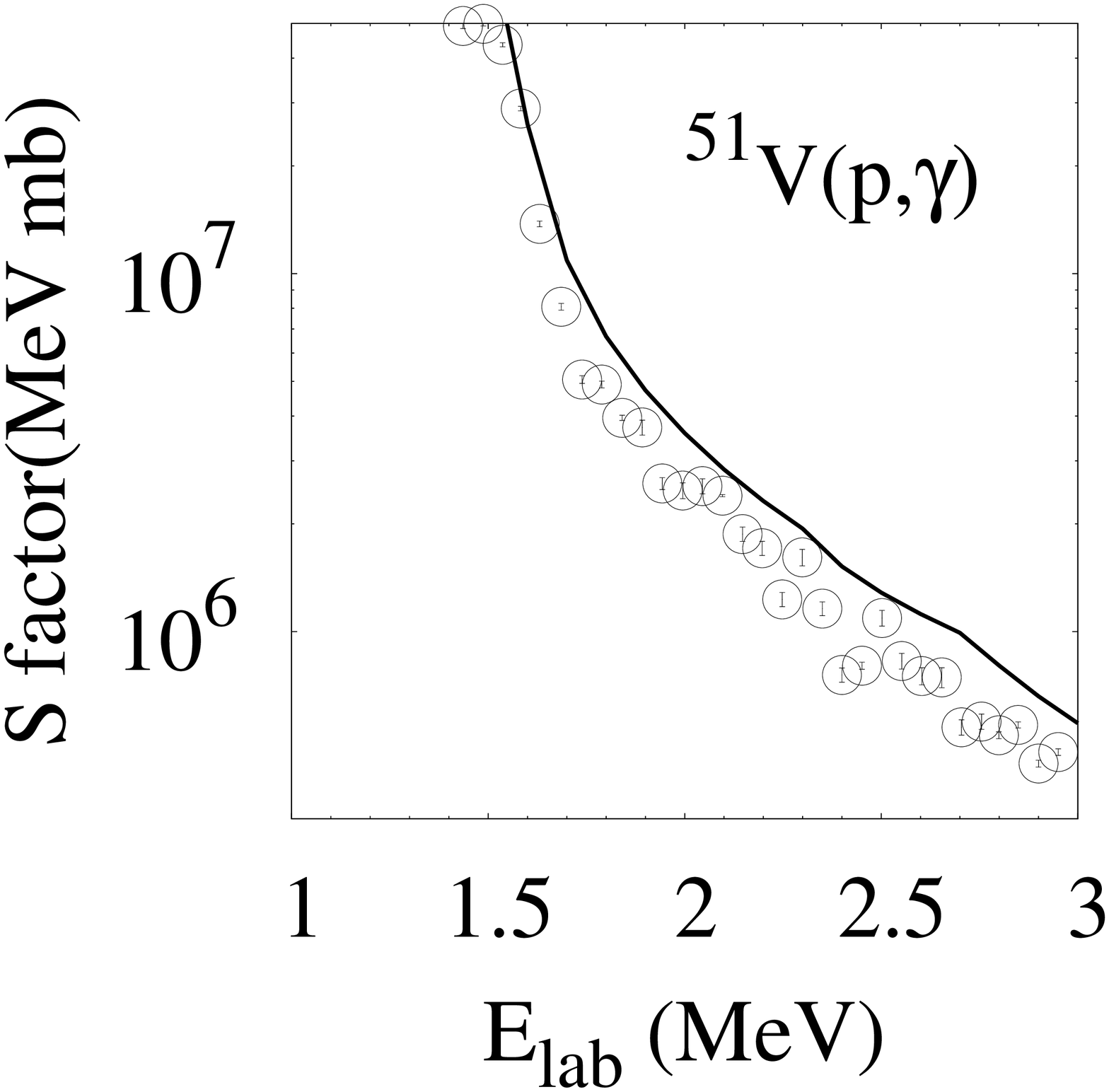}   \\
\multicolumn{2}{c}{\includegraphics[width=.25\textwidth]{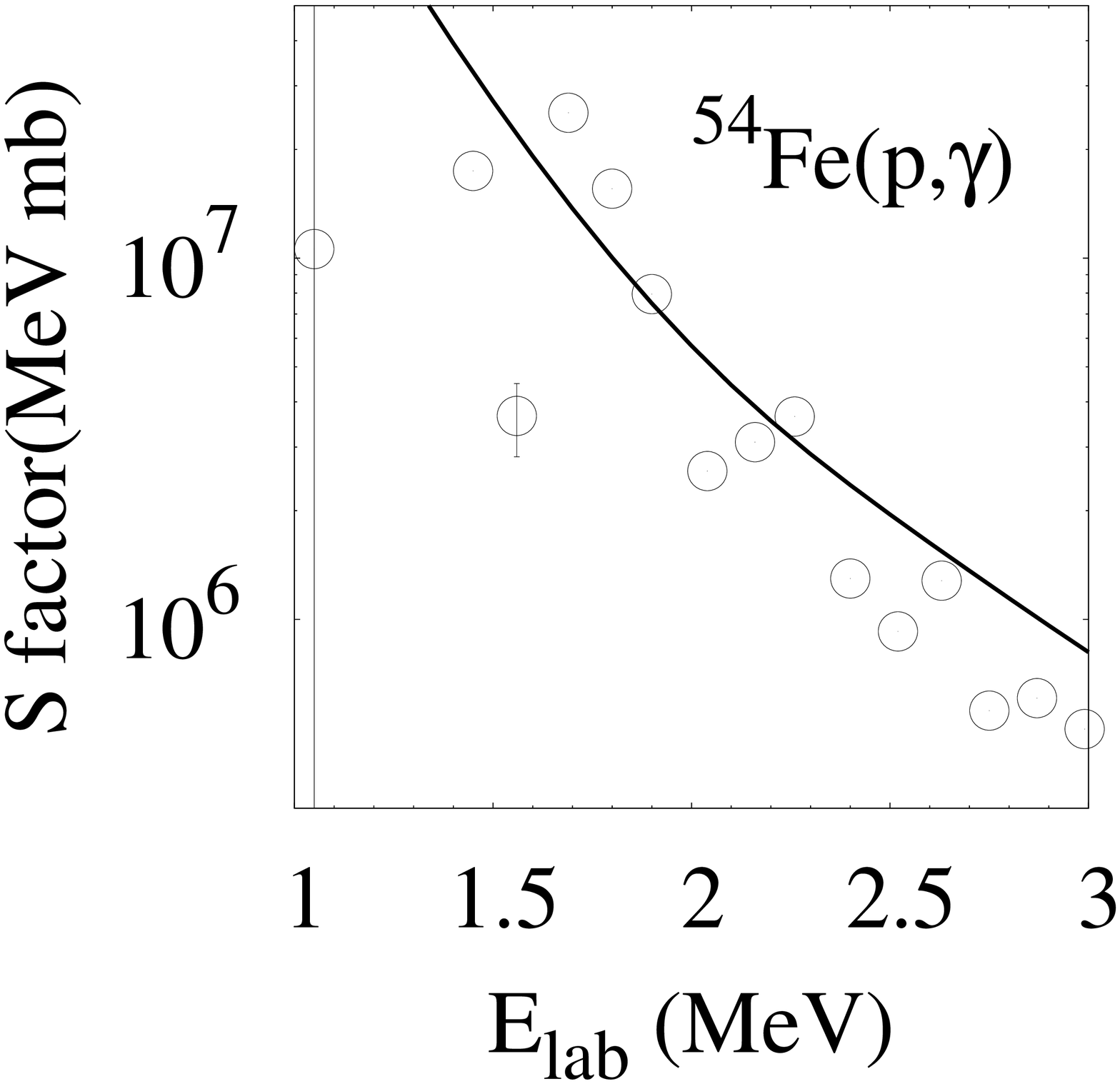}}
\end{tabular}
\caption{\label{fig2}Comparison of astrophysical S factor with available experimental values for the Potassium,Vanadium and Iron isotopes. Solid lines denote our present calculation and the discrete points denote experimental values. Figure specification will be same afterwards unless otherwise mentioned. In most of the cases 
errors, when available, are smaller than the dimension of the points.}
\end{figure}

Proton capture cross section data are available for a large number of targets 
in this mass region. We have calculated the values using our approach and then we have compared those with experimental measurements.
Cross section for the $^{41}$K$(p,\gamma)^{42}$Ca reaction has been measured using
Ge(Li) detectors  of volume 125 cm$^{3}$ below the neutron threshold and 
60 cm$^{3}$ above the neutron threshold by Sevior {\em et al.}~\cite{k41}. Theoretical 
cross sections were calculated with the statistical code HAUSER*4~\cite{hauser4}.
Reasonable agreement with experimental values was obtained by reducing the 
imaginary well depths in global parameters. Our calculation excellently 
reproduces the measurement as can be seen from Fig. \ref{fig2} without any
further modification.

The only stable isotope of vanadium is $^{51}$V. 
The s factors for $(p,\gamma)$ reaction fall steeply above 1.5 MeV of energy due to low 
threshold of $(p,n)$ 
reaction channel ($Q=-1.534$ MeV), a consequence of neutron richness of $^{51}$V.
A competition between these two reaction channels causes a drop in $(p,\gamma)$
cross-section for incident energies just above the neutron threshold. This 
particular low $(p,n)$ threshold in some nuclei are of major importance as it 
provides opportunity to study both reactions within the range of 
astrophysically important bombarding energies.

Zyskind {\em et al.}~\cite{v51} measured the $^{51}$V$(p,\gamma)^{52}$Cr reaction 
cross sections. The $\gamma$ rays and neutrons were detected by a 
73 cm$^{3}$ Ge(Li) detector and BF3 long counter, respectively.
They also compared the average values of the measured cross section 
with two theoretical models, the Kellogg global HF program (KGHFP)~\cite{kghfp}
 and HAUSER*4~\cite{hauser4}. They observed that KGHFP over-predicts the measurement by 
30-50\% while HAUSER*4 exceeds the data by only 2-3\%.
The errors in their measurement arose mainly from target thickness and detector efficiency resulting overall uncertainty of  about 20\%.
The present theory overestimates the measurement by a factor $\sim$ 2 as 
can be seen from Fig. \ref{fig2}. 

The Fe isotope with mass number A=54 is an even-even nucleus with magic neutron 
number. Kennet {\em et al.}~\cite{fe54} measured the  cross section of the reaction 
$^{54}$Fe$(p,\gamma)^{55}$Co using  a 125 cm$^{3}$ Ge(Li) detector in the 
energy range 1.05-3.69 MeV. 
Those cross sections were obtained by the authors 
after summing up three major transitions those resulted 59$\pm$3\% of the total
reaction strength corresponding to
the ground state transitions from first, second and third excited states. 
The absolute error in their measurement was reported to be of the order of 12\%. They carried
out a
statistical calculation using HAUSER*4~\cite{hauser4}  
which overestimated the cross section. In search of an improved fit they 
proposed a new prescription by considering proton imaginary
well depth as a free parameter and obtained a best fit linear relationship between 
the reduced proton imaginary well depth and proton number. 
However, the experimental data was taken at large energy intervals and
 are highly scattered making the comparison difficult. Nevertheless, the 
agreement with our 
theoretical prediction for this case, if not excellent, is within satisfactory 
limit as can be seen from the Fig. \ref{fig2}. The value corresponding to the 
lowest energy has a large error. 
\begin{figure}[htb]
\center
\begin{tabular}{@{}cc@{}}
\includegraphics[width=.25\textwidth]{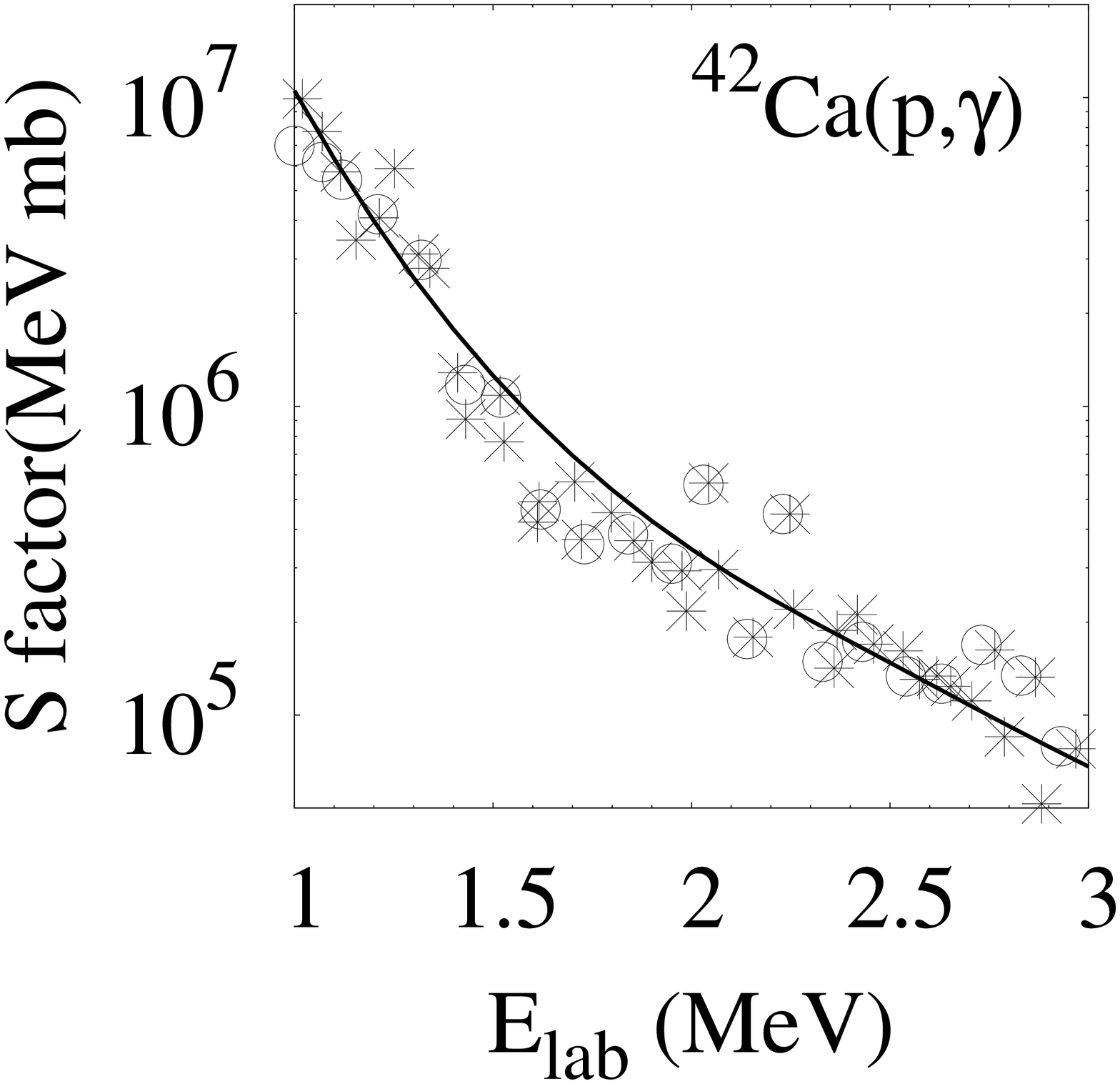} &
\includegraphics[width=.25\textwidth]{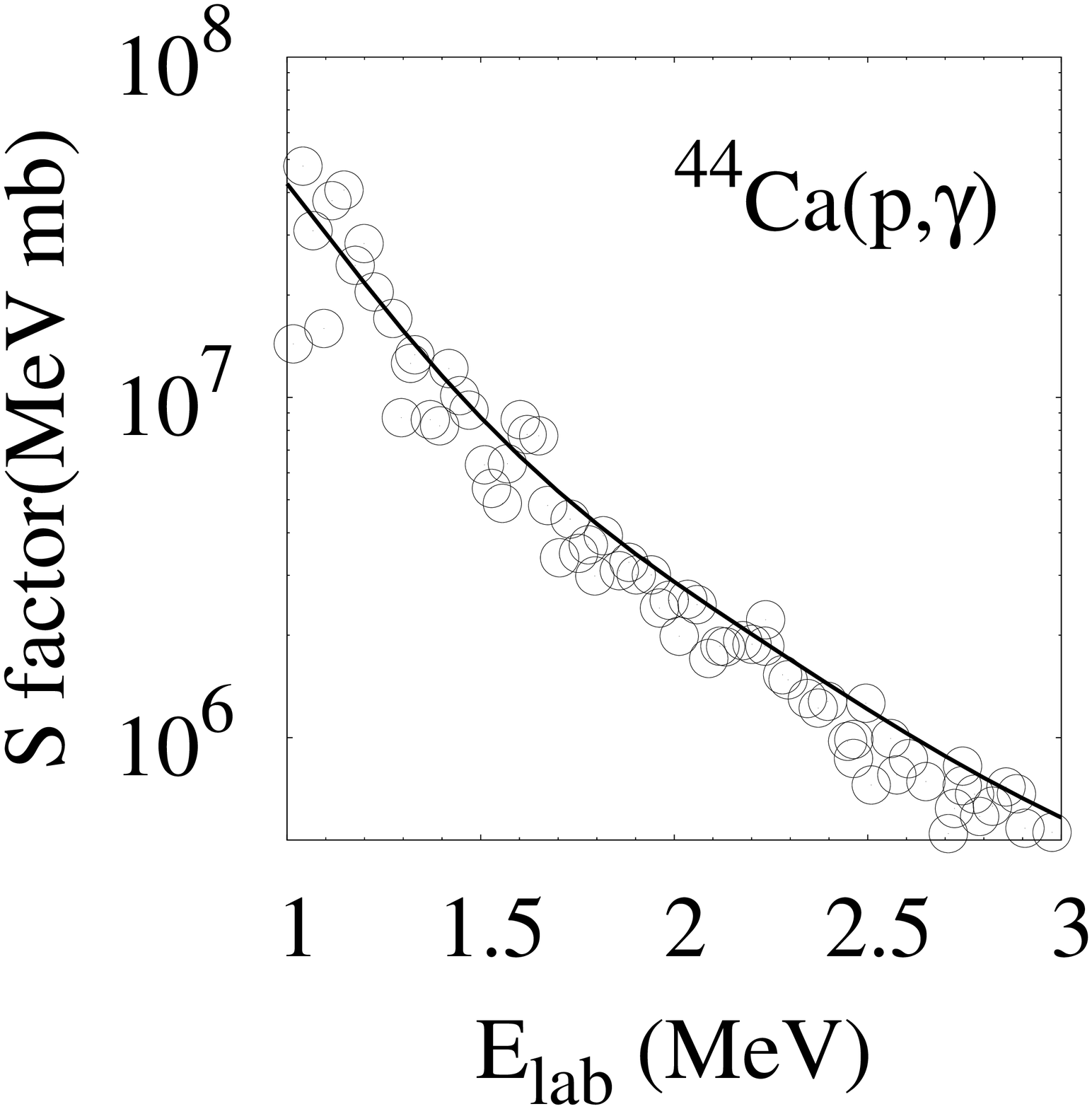}   \\
\multicolumn{2}{c}{\includegraphics[width=.25\textwidth]{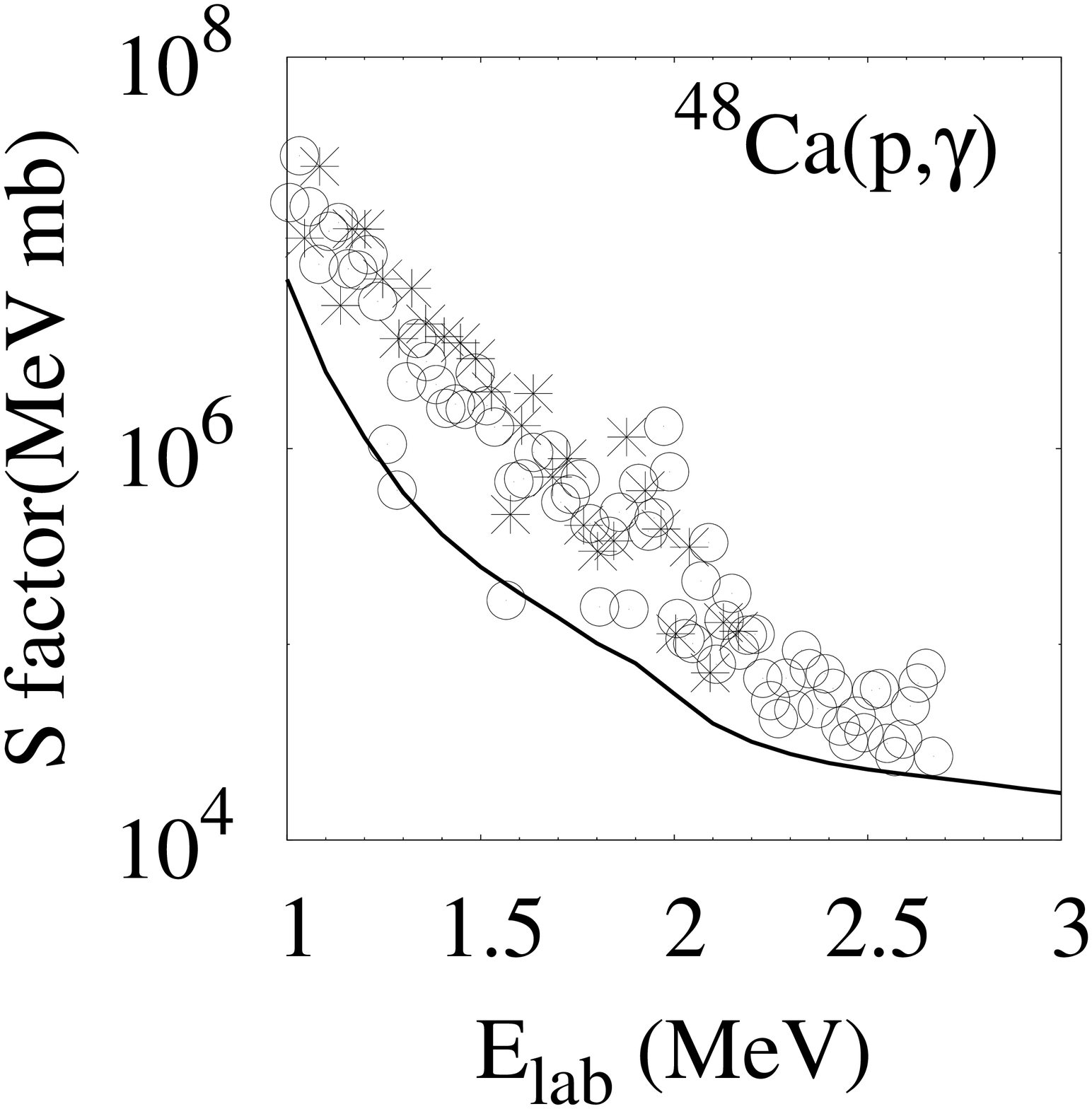}}
\end{tabular}
\caption{\label{fig3}Verification of astrophysical S factor with available experimental values for Calcium isotopes.}
\end{figure}

One of the stable even-even isotopes of calcium is $^{42}$Ca originated mainly
via the explosive oxygen burning in appropriate stellar environment. This is of some importance in the study of  quasi equilibrium achieved 
during explosive silicon burning and takes a very important role in bridging the chain reactions involving masses 28$\leq$A$\leq$45 with that involving masses 45$\leq$A$\leq$62~\cite{woosley}.
It is also a possible contributor to the s-process abundance as it lies along 
the s-process nucleosynthesis path. Proton capture on $^{42}$Ca isotope produces
the relatively long-lived $^{43}$Sc isotope. 

Vlieks {\em et al.}~\cite{ca42a} determined the 
cross-section by measuring annihilation photons emitted after positron decay 
of $^{43}$Sc (half-life=3.89 hour). They compared their measurement with 
theoretical calculation and found good agreement below 4.5 MeV. The
uncertainties in their measurement were mainly from target thickness resulting 
overall uncertainty of 21\% below 3 MeV of energy which further got 
increased to 24\% above 3 MeV energy.

Later Mitchell {\em et al.}~\cite{ca42bca44} measured the same reaction
cross section but with a more direct  technique with 125 cm$^{3}$ Ge(Li) 
detector. 
The total cross section of $^{42}$Ca$(p,\gamma)^{43}$Sc 
was obtained from excitation function of the three major $\gamma$ transitions 
that comprised 65 \% of total strength. They further compared their measurement 
with statistical HAUSER*4~\cite{hauser4} code and found that theoretical prediction 
was higher by a factor of $\sim$ 1.5. They reported the errors associated with
their experimental cross sections for this particular reaction to be 18\%.  Our 
calculation presents  far better agreements with the measurements of both groups than their statistical model calculations,
as can be seen from a comparison of the plots Fig. \ref{fig3} of this work with Fig. 1 of 
Ref.~\cite{ca42a}, and 
Fig. 5 of Ref.~\cite{ca42bca44}.

Mitchell {\em et al.}~\cite{ca42bca44}, in the same work also measured the
 cross section of $^{44}$Ca$(p,\gamma)^{45}$Sc reaction. In the case of $^{44}$Ca,
 yield was determined from the sum of all spectra feeding first excited state and ground state.
 The selected four of all transitions within 3.0 MeV of bombarding energy,
 carrying 68 \% of total strength were used to determine total cross section of this reaction. 
The comparison with HAUSER*4 prediction resulted an overall agreement within a factor $\sim$ 1.3.
 This is more or less similar to the agreement with present theory.

The doubly magic nuclei, $^{48}$Ca has a very long half life against radioactive decay and can be considered as stable for all practical purposes. This nuclei may be produced during explosive carbon burning by reactions of few neutrons and protons with a small admixture of seed nuclei present at the time of star formation~\cite{seed_reaction1}. Being a neutron rich nuclei, it is very important and can be treated as a starting point of production of new nuclei. 
 Experimental data for this reaction has been taken from Ref.~\cite{ca48a} and Ref~\cite{ca48b}.
 The experiment by Kennett {\em et al.}~\cite{ca48a} was carried out using  a 60 cm$^{3}$ Ge(Li) detector with 
detection efficiency determined in situ. The total cross section was
 determined from the 83$\pm$5\% contribution of two transitions to ground 
state from first and second excited states.
The strength was determined from branching ratio averaging over resonances.
Zyskind {\em et al.}~\cite{ca48b} also measured the $(p,\gamma)$ reaction cross section on $^{48}$Ca using a 73 cm$^{3}$ Ge(Li) detector.
The data, even after smoothing, contains considerable fluctuations.

The difference between our theoretical prediction and the experimental data, possibly be attributed to the effect of doubly magic core of the $^{48}$Ca nucleus. The resonances are few and the level density of the compound nucleus system is quite low also. So, the experimental data contain considerable fluctuations and in such a case, statistical model is not expected to be very much effective as HF calculation assumes average over many resonances. Another important aspect is that due to the low lying states of the residual nucleus of competing $(p,n)$ channel, $(p,\gamma)$ cross section changes dramatically. 

\begin{figure}[htb]
\center
\begin{tabular}{@{}cc@{}}
\includegraphics[width=.25\textwidth]{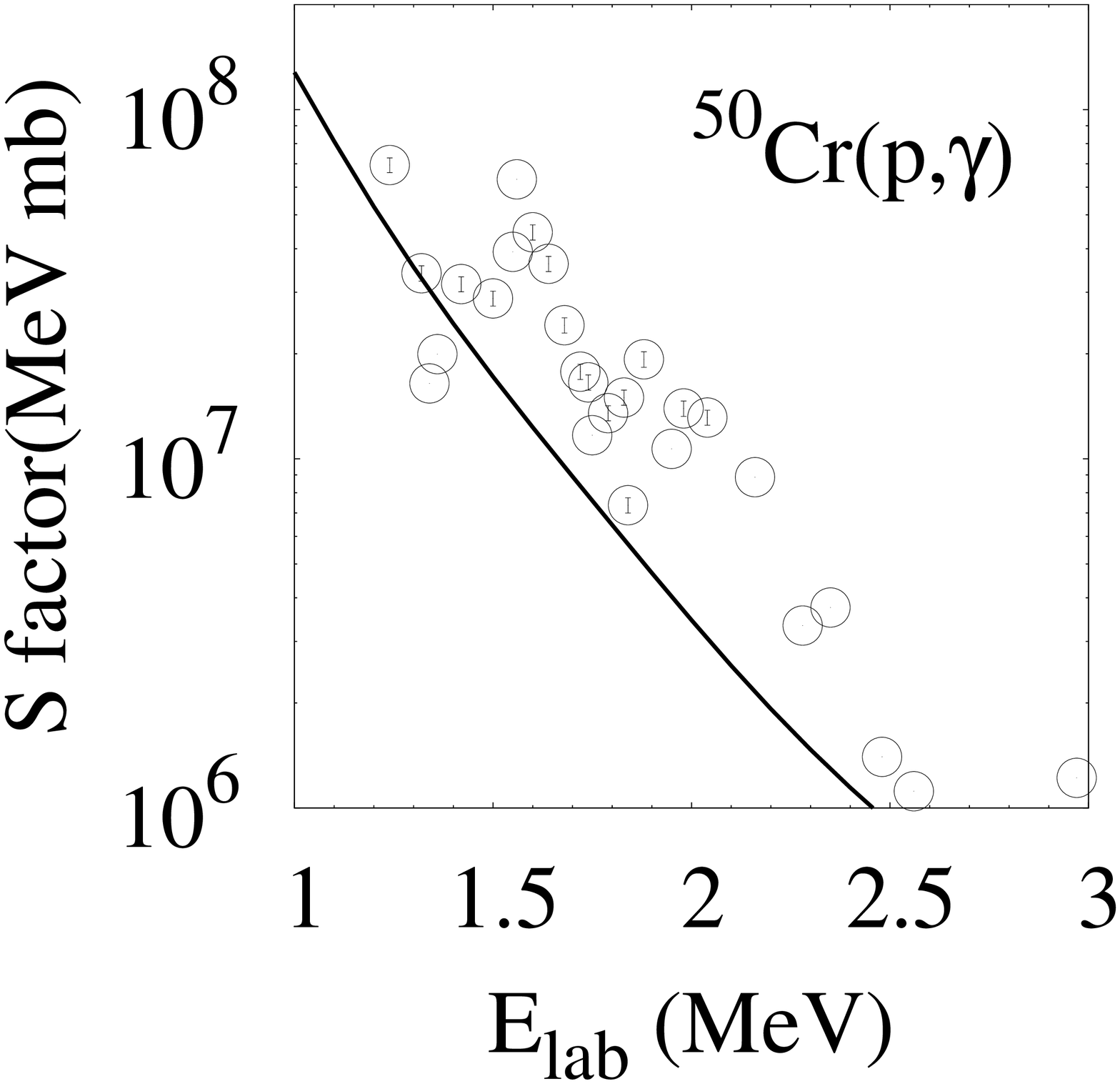} &
\includegraphics[width=.25\textwidth]{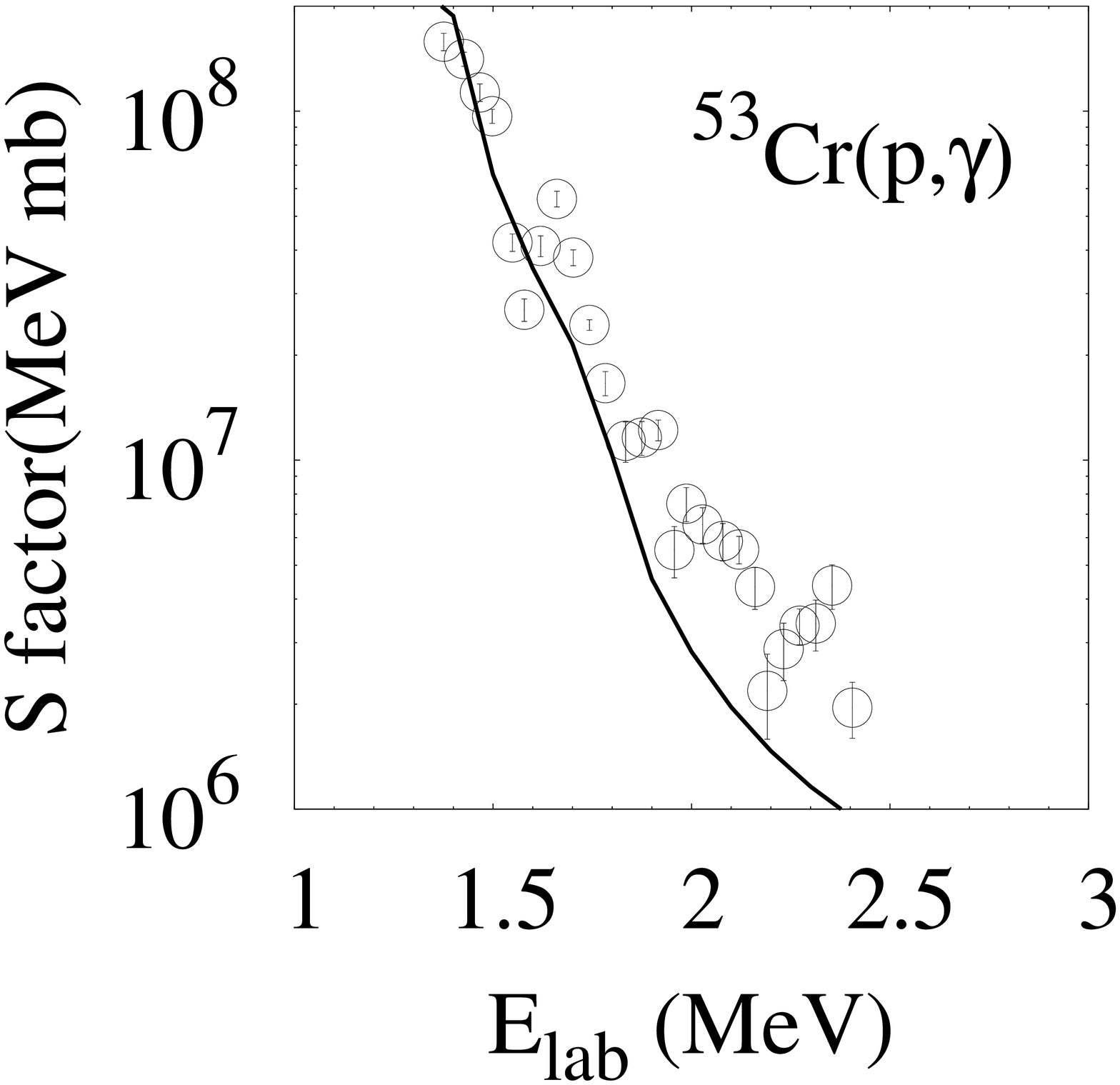}   \\
\multicolumn{2}{c}{\includegraphics[width=.25\textwidth]{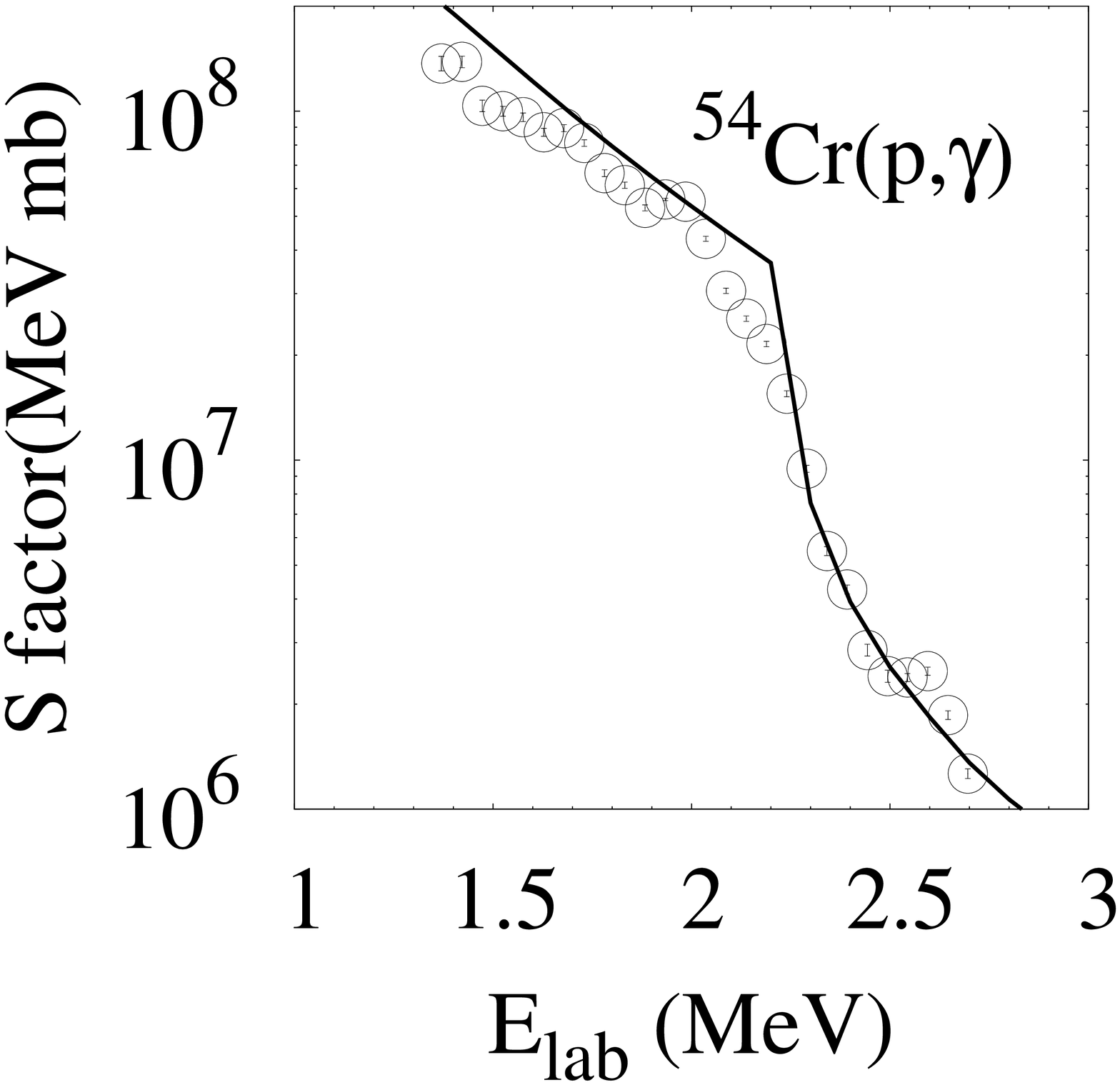}}
\end{tabular}
\caption{\label{fig4}Verification of astrophysical S factor with available experimental values for Chromium isotopes.}
\end{figure}

The experimental data for $(p,\gamma)$ reaction on $^{50}$Cr are taken from
 Krivonosov {\em et al.}~\cite{cr50}. The  
$\gamma$ rays and and $\beta^{+}$ radiation were detected in coincidence.
 They determined the cross sections  by  analyzing principal $\gamma$ rays,
 annihilation radiation as well as $\beta^{+}$ radiation. There are large 
fluctuations in the experimental data which is rather old too. The present 
calculation underestimates 
the average data by a factor $\sim$ 4-5. This, though not very good, is perhaps
reasonable for such a work which aims to set a definite set of normalization 
over entire mass range. A local tuning of parameters can of course provide more 
accurate results.


\begin{figure}[htb]
\center
\begin{tabular}{@{}cc@{}}
\includegraphics[width=.25\textwidth]{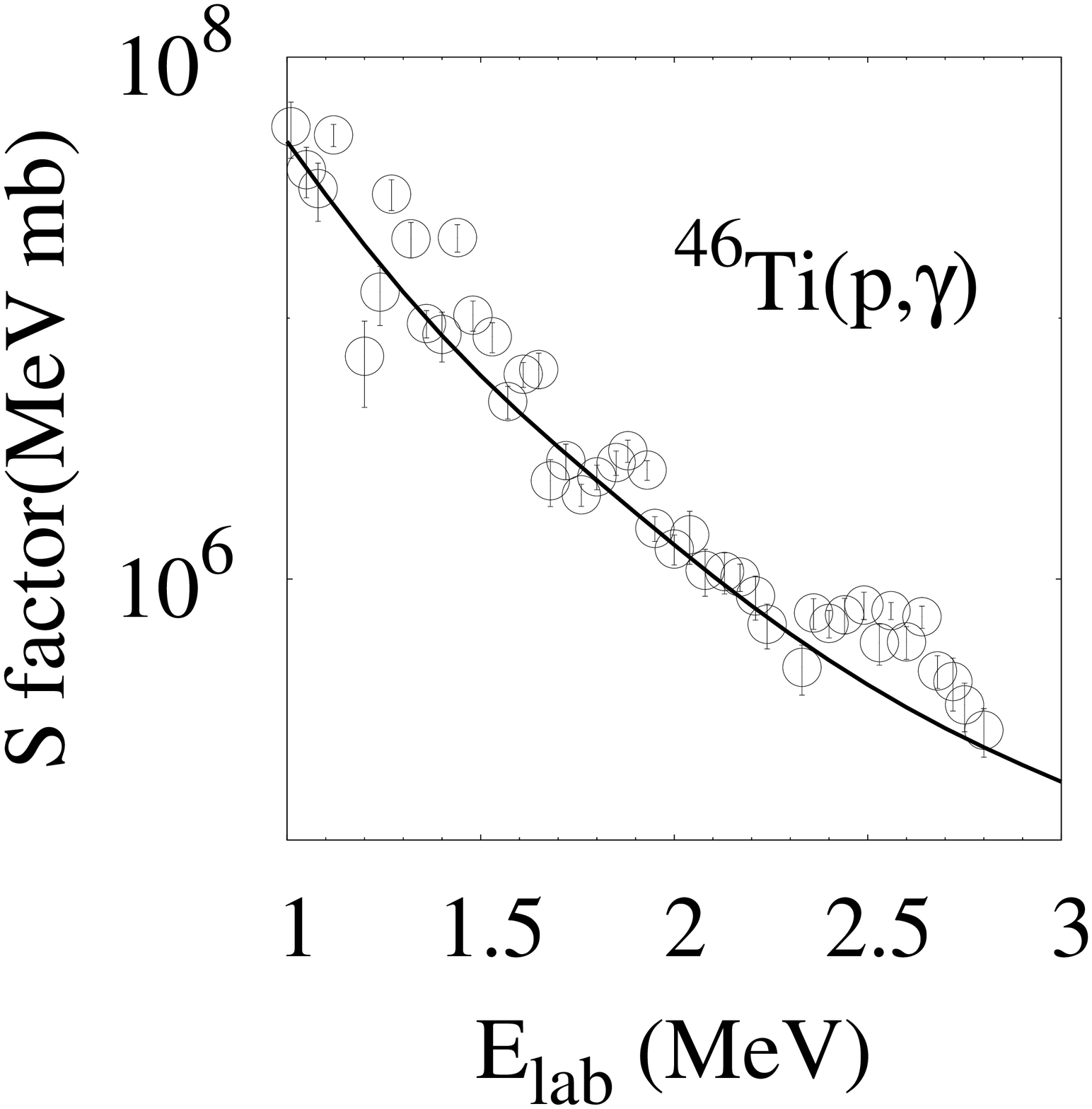} &
\includegraphics[width=.25\textwidth]{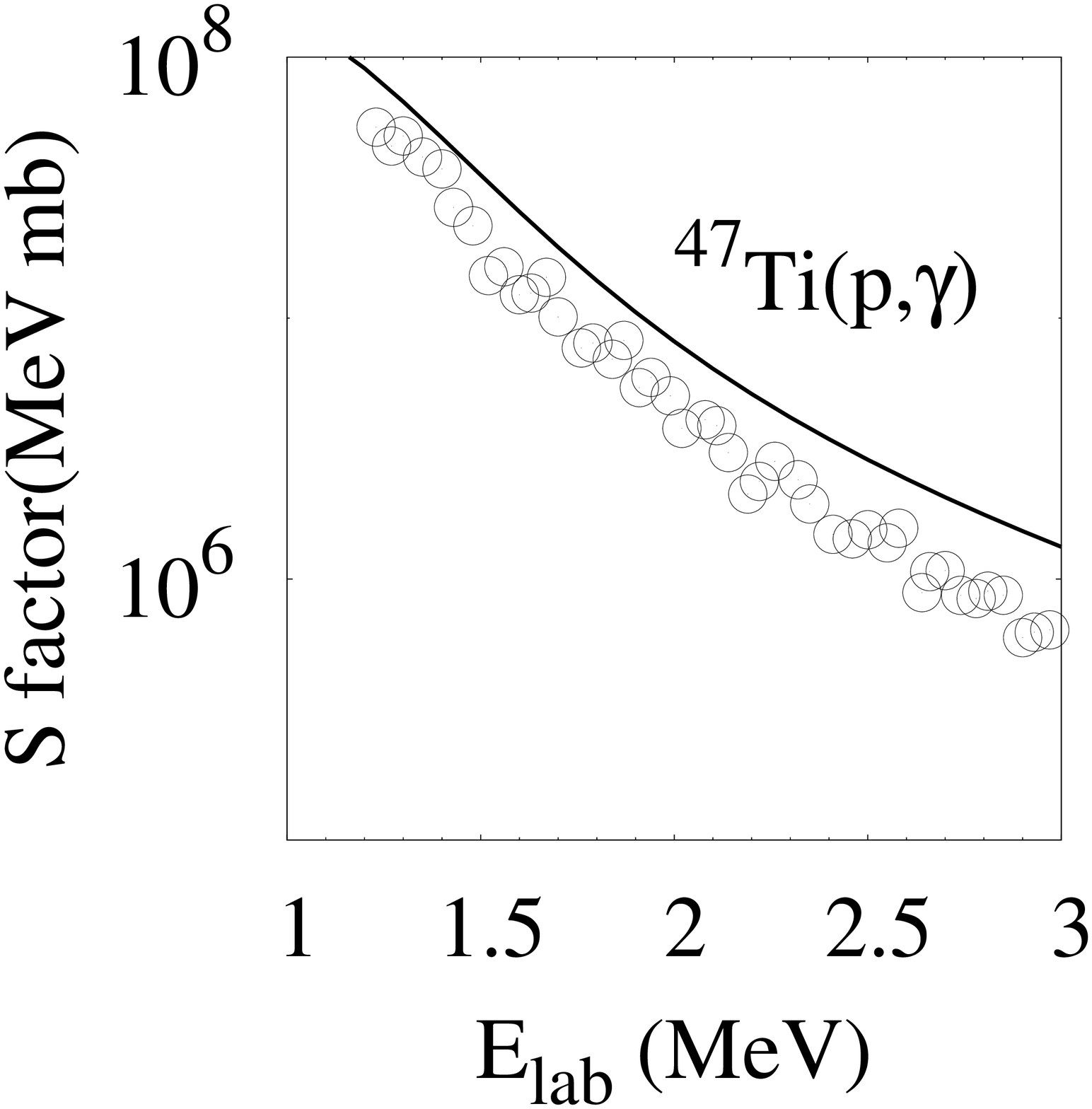}   \\
\includegraphics[width=.25\textwidth]{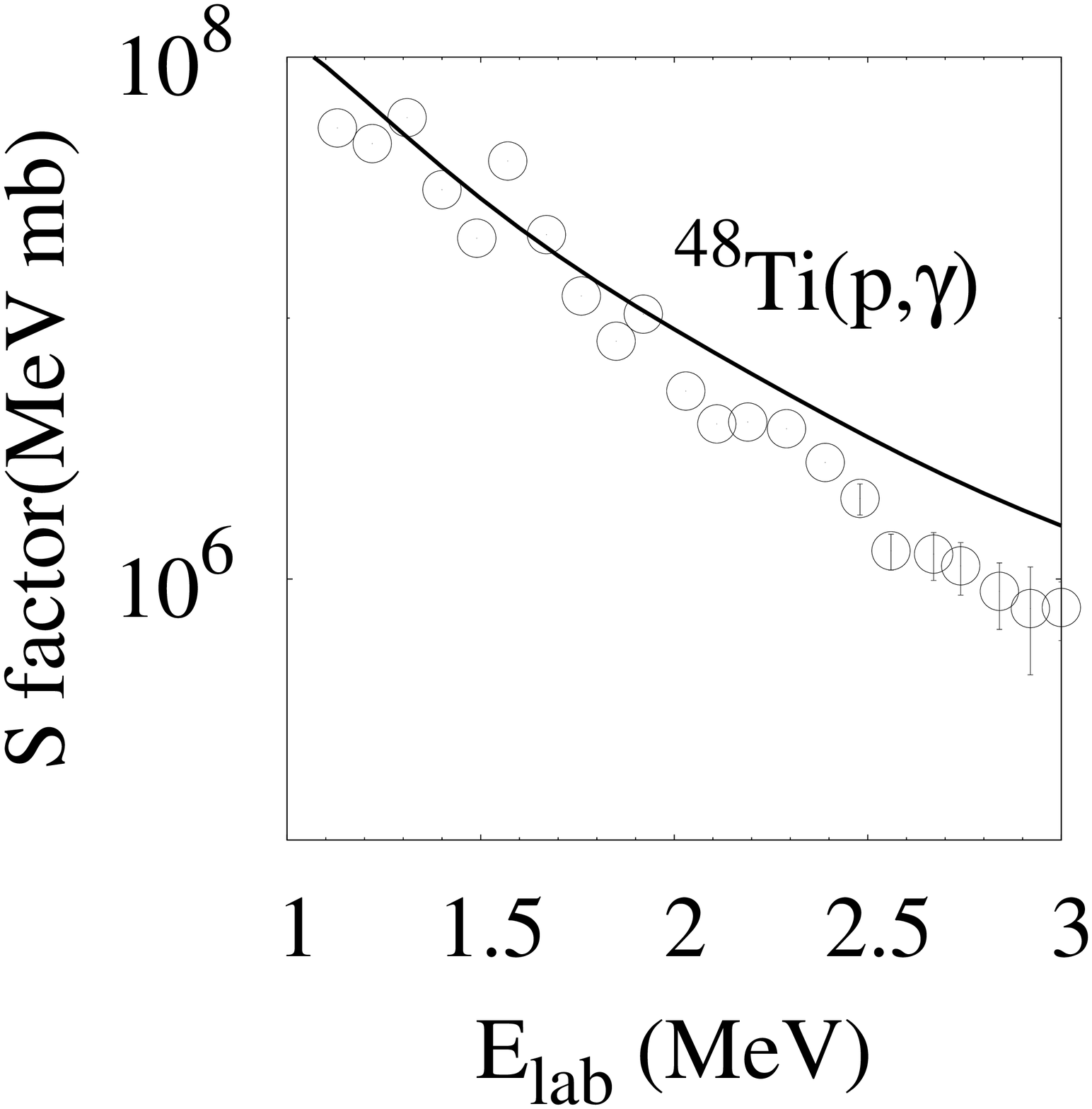} &
\includegraphics[width=.25\textwidth]{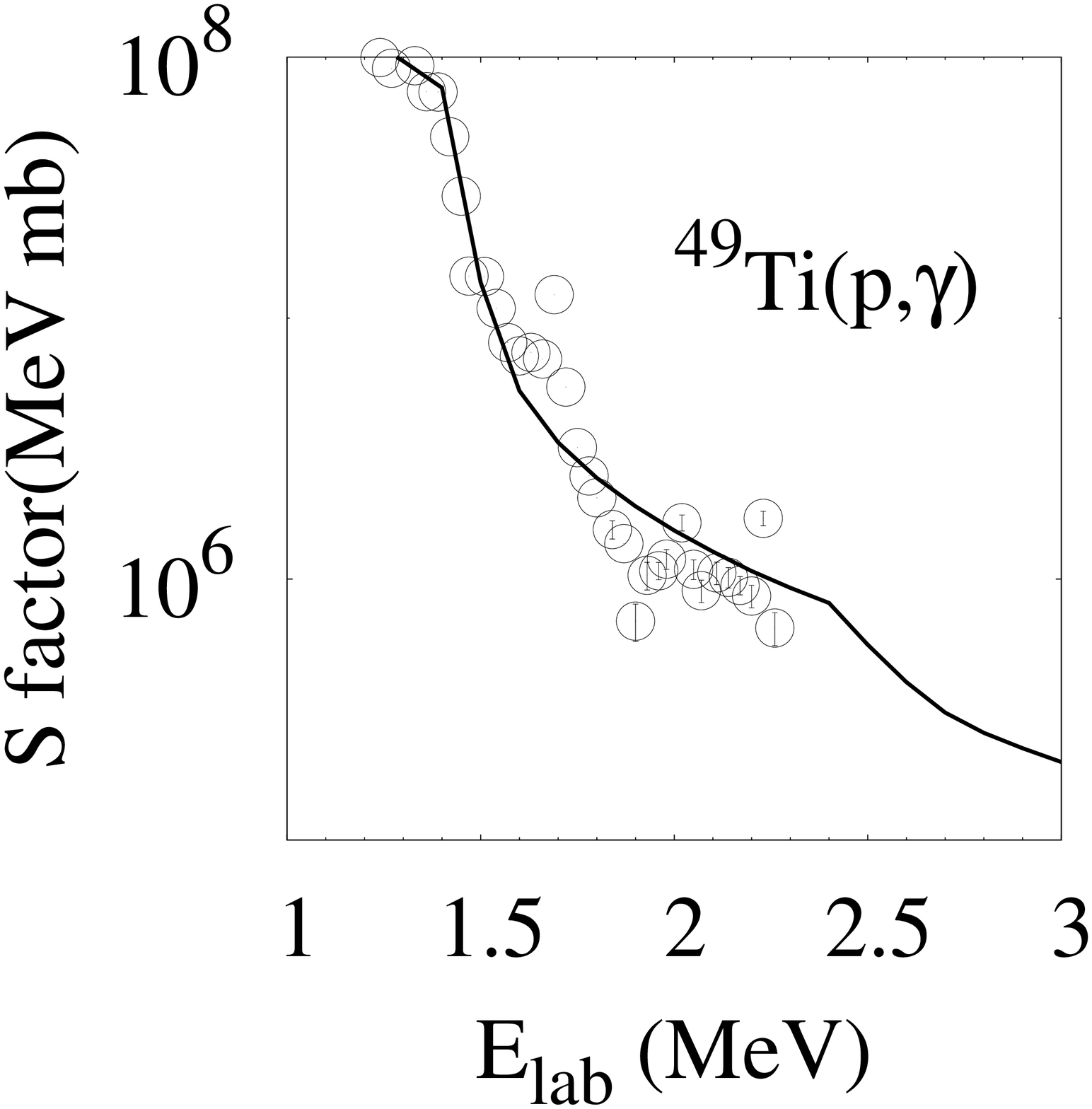}   \\
\multicolumn{2}{c}{\includegraphics[width=.25\textwidth]{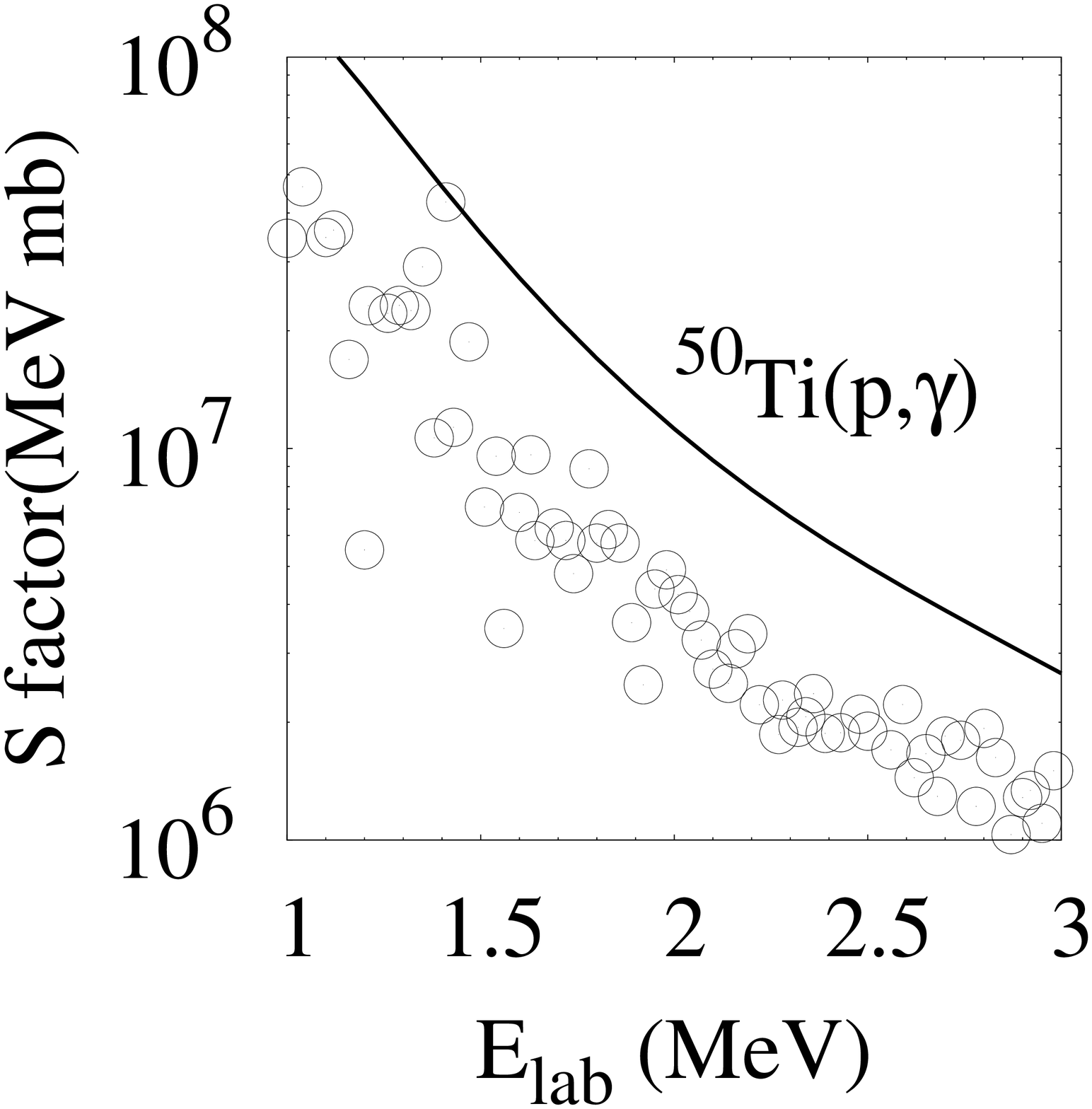}}
\end{tabular}
\caption{\label{fig5}Verification of astrophysical S factor with available experimental values for Titanium isotopes.}
\end{figure}
\begin {figure*}[htb]
\includegraphics[scale=.24]{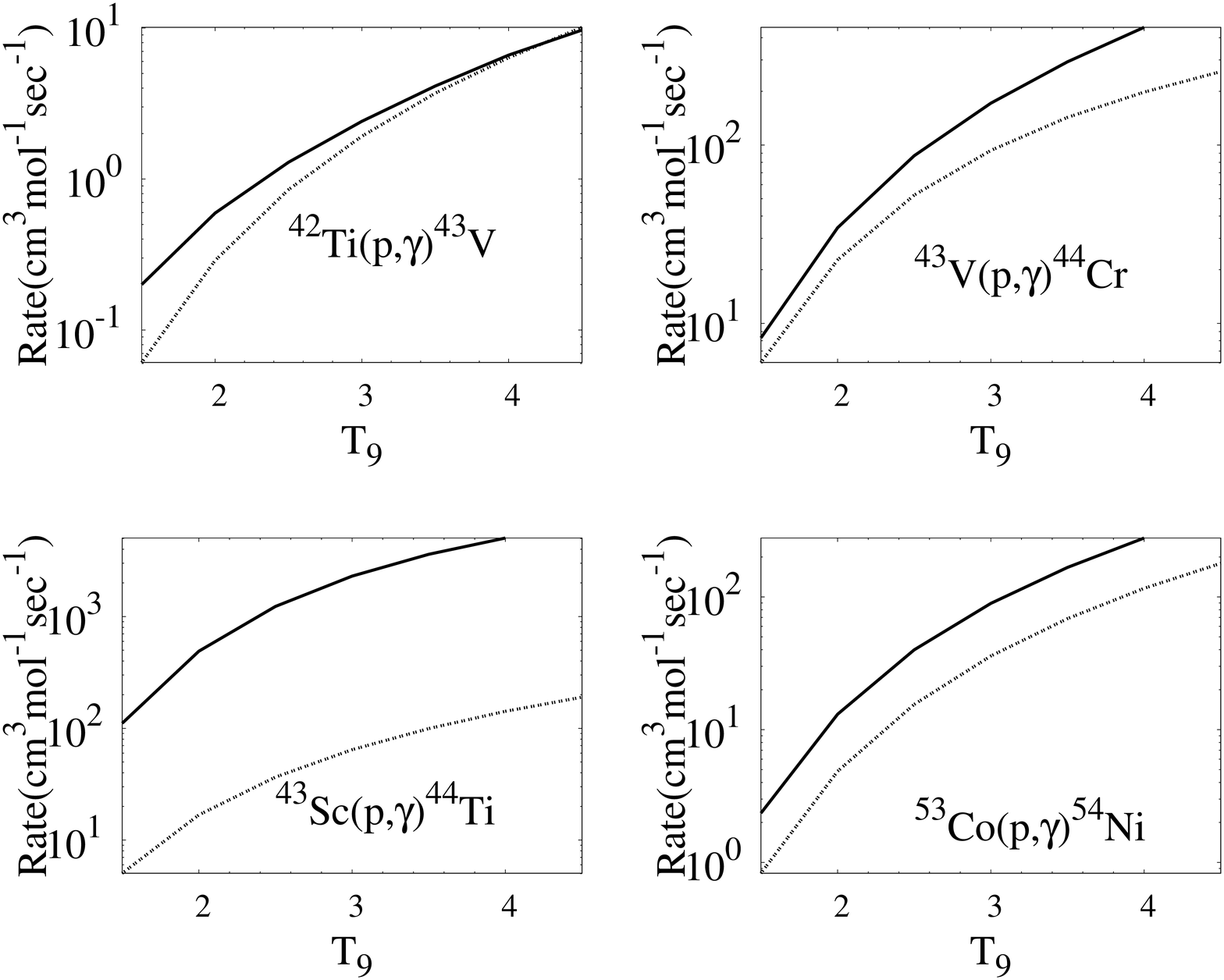}~~~~
\includegraphics[scale=.24]{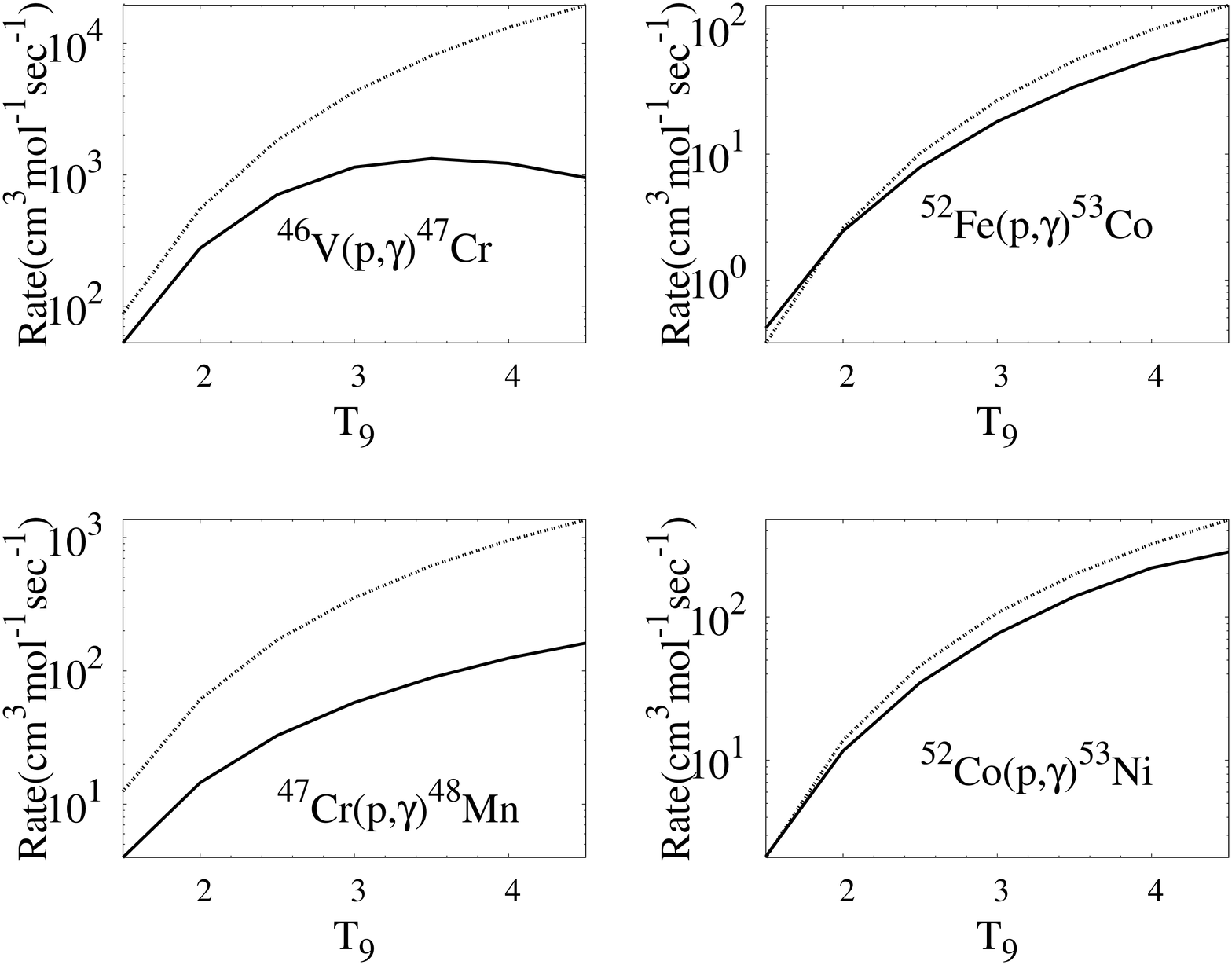}~~~~
\caption{Comparison of proton capture reaction rate with NON-SMOKER result. Solid line denote the rates predicted by our calculation and dotted curve denote the NON-SMOKER result.}\label{rate}
\end{figure*}

Gardner {\em et al.}~\cite{cr53} measured the proton capture reactions on $^{53}$Cr. 
The $\gamma$ rays were detected with a 68 cm$^{3}$ Ge(Li) detector  and the 
analysis was performed from the five transitions
feeding first excited state (0.054 MeV) and ground state of $^{54}$Mn. The 
overall accuracy of the measurement was estimated to be 12\% with the greatest 
source of the error resulted from target thickness uncertainty. Statistical 
model calculations done by the authors using global optical model parameters 
showed satisfactory agreement with 
the experiment. 
Our microscopic theory, however, gives a far better agreement with their 
measurement,
 especially below the $(p,n)$ threshold (at 1.40 MeV) where only $(p,\gamma)$ reaction
 channel is open. 

The experimental data for $(p,\gamma)$ reaction on $^{54}$Cr target is from 
Ref.~\cite{cr54}. A 73 cm$^{3}$ Ge(Li) detector was used to measure the absolute 
cross section of the reaction. The systematic error was reported to be 
$\sim$ 20\%. The experimental excitation function was plotted by the authors 
along with the results of statistical HF calculations of Ref~\cite{fowler} and the results
from the Hauser*4~\cite{hauser4} code.
They found excellent agreement with Ref~\cite{fowler} however the results from the Hauser*4 code showed
some inconsistencies. 
We get excellent agreement with the experimental results perhaps better than both the calculations mentioned above.

Some of the Ti isotopes are important in astrophysics. For example, 
the most strongly deformed even-even Ti nuclei is $^{46}$Ti. 
It is mainly produced during stellar burning stages in AGB stars via rapid 
capture of protons on Sc isotope. Some traces of it is also  observed in some 
collapsing stars, supernovae etc. The $^{46}$Ti isotope contributes also to the (though small) rp-process flux and to the early burning stages of the X-ray bursts affecting the luminosity of the objects. The $^{46,47}$Ti isotopes carry a dominant flow of nucleosynthesis from mass 45 bottle neck to the elements of iron group~\cite{woosley}. 

The experimental cross section values for $^{46}$Ti$(p,\gamma)^{47}$V, $^{47}$Ti$(p,\gamma)^{48}$V, $^{48}$Ti$(p,\gamma)^{49}$V
 reactions are taken from Ref.~\cite{ti464748}. The $(p,\gamma)$ reaction yields were measured with a 125 cm$^{3}$ Ge(Li) detector
 with efficiencies accurate within $\pm$ 5\%. However for cross section above neutron threshold 
on $^{47}$Ti a 60cm$^{3}$ Ge(Li) detector was used. They determined the 
cross-sections from the selected transitions
after summing up all spectra over a wide energy range and correcting for the 
fraction of intensity belonging to the selected lines.
The data 
were compared with HAUSER*4 code calculations and agreement were
within 30\%, 50\%, and 20\% for $^{46}$Ti, $^{47}$Ti, and $^{48}$Ti,
respectively as reported by the authors. The authors concluded that the
agreement was good for globally parameterized statistical code. Our theory 
reproduces the measurements 
reasonably well, within a factor of $\sim$ 1.5 or less, for all three cases. However, in case of $^{46}$Ti, our result varies $\sim$ 10 with another set of experimental data avilable in Ref.~\cite{famiano} and the reason of this discrepancy is not very clear. 

Kennet {\em et al.}~\cite{ti49} measured the cross section for the $^{49}$Ti$(p,\gamma)^{50}$V reaction.
The $(p,\gamma)$ excitation function was obtained by observing the transition 
from the first excited state to the ground state of 
$^{50}$V which is reported to carry 97\% of the total
 $(p,\gamma)$ strength. 
The authors  compared their results with the calculation from
the HAUSER*4 code~\cite{hauser4} and found agreement within 30\%. 
The present model
gives an excellent agreement with experiment for this particular reaction.

The even-even nucleus $^{50}$Ti contains magic number of neutrons and is the 
most neutron rich among the five stable isotopes of titanium.  
The experimental data for proton capture on $^{50}$Ti is from Ref.~\cite{ti50}. The experiment was carried out with a Ge(Li) 
detector of dimensions 125 cm$^{3}$ and 60 cm$^{3}$, below and above the neutron threshold, respectively.
A comparison with the results of HAUSER*4 code~\cite{hauser4} revealed that the code 
overestimated the proton transmission coefficient at the entrance channel 
resulting in overpredictions 
by a factor of $\sim$ 3. Our results also show significant overestimation in 
the s-factor for this reaction. This can perhaps be attributed to the limitation
of the statistical model near a closed shell where large shell gaps lead to
low level densities. 

It is worth noting that the data for radiative proton capture reactions
in astrophysical energy range are in scarce.
Even most of the existing data are also
very old and lacks the application of modern techniques. Thus possibility of 
the presence of errors
associated with the data is high. In most of the cases 
individual error associated with each data point is not available. Hence,
it is extremely difficult to use these data with much 
reliability.  Our aim is to present a unique set of parameterization over the 
entire mass range so that the cross section can be extrapolated to those targets for which the same is unavailable till the date. We do not expect that our results will need to be modified after any remeasurement of the values.

\subsection{Astrophysical $(p,\gamma)$ reaction rate calculation}
Inside stars, nuclides not only exist in their ground states but also in their excited states and a thermodynamic equilibrium holds to a very good approximation. The assumption of a thermodynamic equilibrium combined with the compound nucleus cross sections for the various excited states then allows us to produce Maxwellian averaged reaction rates, which are important inputs for stellar evolution models. In astrophysical environment e.g. like x-ray bursters, the relative populations of levels of target nuclei obey Maxwell-Boltzmann distribution. The effective stellar rate in the entrance channel at temperature $T$ taking due account of the contributions of the various target excited states is expressed as
\begin {equation}
\begin{aligned}
N_{A}<\sigma v>_{\alpha\alpha^{\prime}}^{*}(T)=(\frac{8}{\pi \mu})^{1/2}\frac{N_A}{(kT)^{3/2}G(T)}\times\\\int_{0}^{\infty} \sum_{\nu}\frac{(2I^{\nu}+1)}{(2I^{0}+1)}\times \sigma^{\nu}_{\alpha\alpha^{\prime}}(E) E \exp(-\frac{E+E^{\nu}_{x}}{kT})dE
\end{aligned}
\end{equation}
where $\nu$ represents various excited states in the nucleus and
\begin{equation}
G(T)=\sum_{\nu}{ \frac{(2I^{\nu}+1)}{(2I^{0}+1)}\exp{-(\frac{E^{\nu}_{x}}{kT})}  } 
\end{equation}
is the $T$-dependent normalized partition function.   
Fig.\ref{rate} shows the reaction rates for $(p,\gamma)$ reaction from the 
present calculation being compared with the rates from the NON-SMOKER 
code~\cite{nonsmoker,nonsmoker_web}. NON-SMOKER code is an improvisation 
of the well-known reaction code SMOKER~\cite{smoker} with modified level 
density description,
explicit isospin mixing treatment, width fluctuation correction, GDR energies 
and widths. The NON-SMOKER code uses a HF calculation based on masses from Finite Range Droplet Model (FRDM)~\cite{frdm}.
We have plotted the proton capture rates in the temperature range of 1 GK to
4.5 GK  for some reactions in the mass region in Fig. \ref{rate} in the cases where our results differ significantly from the NON-SMOKER results. The numerical values of the reaction rates are given in the Supplemental Material~\cite{sm}. 
As can be seen that the reaction rates determined from our theory is more than
the NON-SMOKER rates in the cases of $^{42}$Ti, $^{43}$V,
 $^{43}$Sc, $^{53}$Co targets. Our calculated values are smaller than the
NON-SMOKER rates in rest of the cases.
In the case of $^{52}$Fe the present rate exceeds NON-SMOKER prediction above 
2 GK and below this temperature it is less. On the other hand, our present calculation of astrophysical rates for the $(p,\gamma)$ reactions of $^{43}$V and $^{52}$Co, got merged with the NON-SMOKER calculation  up to 2 and 1.5 GK respectively. The difference between the two calculated rates increases with temperature in case of $^{46}$V$(p,\gamma)$ reaction. It will be very interesting to see the effect of these rates in the abundance calculation of nuclei in relevant astrophysical environment.

\section{Summary}

To summarize, we calculated the cross sections for $(p,\gamma)$ reactions in the mass range 40-54
in the relevant Gamow energy window appropriate for low energy astrophysical 
environment  using the  well known reaction code TALYS1.6. Charge radii and binding energy values of various stable nuclei calculated using RMFT have been compared in this concerned mass region A = 40-54 with the available experimental data. The DDM3Y  NN interaction is folded  
with target nuclear densities calculated from RMFT to construct the optical potential which is needed for HF statistical model calculation and after proper normalization it has been used to verify the theory with the observed experimental data. The $(p,\gamma)$ reaction rates are calculated and plotted along with NON-SMOKER reaction rates. The main feature of our work is to place all the nuclei considered in this mass region A = 40-54 at the same footing and to use same methodology for all of them to avoid systematic error. 

\section{Acknowledgement}
The authors acknowledge the financial support provided by University Grants Commission, Department of 
Science and Technology, Alexander Von Humboldt Foundation, and the University of Calcutta.

\end{document}